# Apparent Positions of Planets

Chol Jun Kim,  Kyong Il Pak, Sin Chol Hwang, Chol Min Choe, Jin Hyok Choe, Ui Ri Mun

*Department of Physics, **Kim Il Sung** University,*

*RyongNam Dong, TaeSong District, Pyongyang, DPR Korea*

**Abstract**

The apparent positions of planets are determined by means of the fundamental ephemerides, the precession-nutation models of the Earth, the gravitational effects and aberrations et al.
Around 2000, many astrometrical conceptions, models and theories had been newly defined and updated: for the fiducial celestial reference system, the ICRS is introduced, the fundamental ephemerides - DE405/LE405 et al., precession-nutation model – IAU 2000A/IAU 2006 model.
Using the traditional algorithm and the updated models, we develop the system of calculating the apparent positions of planets. The results are compared with the "Astronomical Almanac" and proved in their correctness.

**Introduction**

The apparent position of a celestial object is related to the true equator and equinox of date and its definition needs taking into account of the fundamental ephemerides of solar system objects, the precession-nutation of the Earth, the gravitational effects and aberrations et al. The fundamental ephemerides give the position of solar system objects with respect to the fiducial system like previously used the equator and equinox of the epoch. The Earth precession and nutation models determine the transformation between reference systems of different dates due to the precession and nutation of the Earth.

Just before and after 2000 many astrometrical conceptions and models had newly been defined and updated. In 1997 International Astronomical Union (IAU) had defined ICRS and ICRF based on the VLBI observations of extragalactic radiosources. International Celestial Reference System (ICRS) is the "space-fixed", kinematically non-rotating reference system which seems to be the inertial system. International Celestial Reference Frame (ICRF), the realization of ICRS, is the list of the positions of radio sources that define the ICRS. In 2000, instead of previous IAU 1976 Precession  Model and IAU 1980 Theory of  Nutation, IAU introduced IAU 2000A precession-nutation model which is also called MHB2000 precession-nutation  model[1]. However IAU 2000A precession used the updated IAU 1976 precession and it is not consistent with IAU 2000A nutation so IAU recommended the development of the dynamically-consistent precession model. In 2006, IAU adopted "IAU 2006" precession-nutation model, which is based on the works of the Working Group on precession and the ecliptic[2]. They developed the new precession model and it is called "P03" model. The IAU 2009 resolutions adopted a new system of astronomical constants and improved realization of the ICRF.

Based on the newly defined celestial reference systems and the International Terrestrial Reference System (ITRS) which is defined by the International Union of Geodesy and Geophysics (IUGG), the new conceptions like earth rotation angle (ERA), Celestial Intermediate Origin (CIO), Celestial Intermediate Pole (CIP) et al. are introduced[3]. The Jet Propulsion Laboratory (JPL) had created the ephemeris DE405/LE405 (or simply DE405) which gives the positions and movements of the solar system objects with respect to ICRF instead of the equator and equinox of the epoch in the previous ephemerides DE1xx and 2xx. After DE 405, all the ephemerides are with respect to ICRF and ICRF2.

The adoption of the new systems, conceptions and models need the new calculation system of positions of the major planets in solar system. We used the traditional calculation algorithm of positions of planets, considered the gravitational effects and replaced the old models with the updated models.

**Algorithm**

Apparent positions of planets are calculated as follows:

1. The moment given in TT should be converted to moment in TDB.

Time argument in ephemerides is TDB (Barycentric Dynamical Time) time which is the successor of previously used $T_{eph}$ (Ephemeris Time). The input argument in Astronomical Almanac, however, is TT time so that in order to get the planetary positions from the ephemerides (e.g., DE405) there needs the conversion between the time scales. The conversion is as follows[4]:

$$\begin{aligned}
T_{eph} \approx \text{TDB} \approx \text{TT} &+ 0.001657 \sin(628.3076\,T + 6.2401) \\
&+ 0.000022 \sin(575.3385\,T + 4.2970) \\
&+ 0.000014 \sin(1256.6152\,T + 6.1969) \\
&+ 0.000005 \sin(606.9777\,T + 4.0212) \\
&+ 0.000005 \sin(52.9691\,T + 0.4444) \\
&+ 0.000002 \sin(21.3299\,T + 5.5431) \\
&+ 0.000010\,T \sin(628.3076\,T + 4.2490) + \ldots
\end{aligned} \qquad (1)$$

where the coefficients are in seconds, the angular arguments are in radians, and $T$ is the number of Julian centuries of TT from J2000.0: $T = (\text{JD(TT)} - 2451545.0)/36525$. In above formula the maximum error between 1600 and 2000 is about 10μs. TDB time and $T_{eph}$ have the equal average rate.

2. For the TDB time, after loading data from ephemerides files (e.g., DE405.bin) and using the Chebyshev polynomials, the barycentric position $\mathbf{E}_B$ and velocity $\dot{\mathbf{E}}_B$ of the Earth are evaluated. And the barycentric positions of a planet $\mathbf{Q}_B$, the Sun - $\mathbf{S}_B$ (with respect to ICRS) are evaluated. Vectors $\mathbf{E}_B, \mathbf{Q}_B, \mathbf{S}_B$ are in au. Vector $\dot{\mathbf{E}}_B$ is in au/day.

3. The light time between the planet and the observer. The above calculated barycentric position of the planet $\mathbf{Q}_B$ and one of the Earth $\mathbf{E}_B$ are for the same TDB moment. However, when the light from the planet arrived at the Earth, the position of the planet has been changed. So the observed position of the planet at given TDB moment is one at the time of the light time ago.

The position of the planet at the time of light time ago is calculated by the iteration method. $\mathbf{Q}_B$ is not the planetary position for the observation at the given TDB moment so that the light time obtained from $\mathbf{Q}_B$ and $\mathbf{E}_B$ does not correspond to the correct observation TDB moment. We can set this light time as zero approximation and calculate the planetary position before this time, then recalculate the light time to observer at this position. This cycle should be repeated until the convergence is achieved. The convergence will reach soon.

In the astronomical yearbook or almanac the apparent positions of planets correspond to the observer at geocenter at TT moment so that we can regard the barycentric position $\mathbf{E}_B$ of the earth (or geocenter) as position of the observer. Then at each steps of iteration we can do following calculations:

For the given TDB moment $t$, the initial light time is given as $c\tau = |\mathbf{Q}_B(t) - \mathbf{E}_B(t)|$. Then

$$\begin{aligned}
\mathbf{E} &= \mathbf{E}_B(t) - \mathbf{S}_B(t) \\
\mathbf{Q} &= \mathbf{Q}_B(t - \tau) - \mathbf{S}_B(t - \tau)
\end{aligned} \qquad (2)$$

are evaluated. The geocentric position of the planet can be obtained:

$$\mathbf{P} = \mathbf{Q}_B(t-\tau) - \mathbf{E}_B(t) \tag{3}$$

Now we can calculate the light time $\tau$:

$$c\tau = P + \frac{2GS}{c^2} \ln \frac{E+P+Q}{E-P+Q} \tag{4}$$

The second term means the relativistic delay of electromagnetic wave in the solar gravitational field. $GS = 1.327\ 124\ 400\ 179\ 87 \cdot 10^{20} \text{m}^3 \cdot \text{s}^{-2}$ stands for the heliocentric gravitational constant. $GS/c^2 = 9.8704 \cdot 10^{-9}$ au is the gravitational radius of the sun, $c$ is light velocity, $E$, $P$, $Q$ are the modules of corresponding vectors. Above evaluations should be repeated until the light time $\tau$ is converged. After $\tau$ is converged we can obtain the correct barycentric position of the planet $\mathbf{Q}_B(t-\tau)$ observed at the given moment.

Now dividing vectors $\mathbf{P}$, $\mathbf{Q}$, $\mathbf{E}$ by modules $P$, $Q$, $E$, we obtain the unit vectors $\mathbf{p}, \mathbf{q}, \mathbf{e}$.

4. We calculate the geocentric direction of the planet $\mathbf{p}_1$, considering the gravitational deflection of light.

$$\mathbf{p}_1 = \mathbf{p} + \frac{2GS}{c^2 E \cdot A} \frac{(\mathbf{p} \cdot \mathbf{q})\mathbf{e} - (\mathbf{e} \cdot \mathbf{p})\mathbf{q}}{1 + \mathbf{q} \cdot \mathbf{e}}, \tag{5}$$

where $\mathbf{p} \cdot \mathbf{q}$, $\mathbf{e} \cdot \mathbf{p}$, $\mathbf{q} \cdot \mathbf{e}$ are scalar products, $A = 1.495\ 978\ 706\ 909\ 8932 \cdot 10^{11}$m is a astronomical unit and converts $E$ from au to meters.

The correction values due to gravitational deflection of light are not essential when the angle between the planet and the sun is great, but in conjunction it reaches 2 arc seconds so that the variation rate of the planetary position is not even.

We consider the light deflection due to not only the sun, but the major planets (e.g. Jupiter), when the geocentric direction of the planet is determined as

$$\mathbf{p}_1 = \mathbf{p} + \frac{2GS}{c^2 E \cdot A \cdot rm} \frac{(\mathbf{p} \cdot \mathbf{q})\mathbf{e} - (\mathbf{e} \cdot \mathbf{p})\mathbf{q}}{1 + \mathbf{q} \cdot \mathbf{e}} \tag{6}$$

, where $rm = M_{sun}/m$ ($m$ is mass of the planet) is the reverse ratio of mass of the planet to mass of the sun, and is named the reverse mass of the planet. The reverse masses of typical planets are:

| | | | |
|---|---|---|---|
| Mercury | 6 023 600 | Jupiter | 1 047.3486 |
| Venus | 408 523.71 | Saturn | 3 497.898 |
| Earth + Moon | 328 900.5614 | Uranus | 22 902.98 |
| Mars | 3 098 708 | Neptune | 19412.24 |
| | | Pluto | 135 200 000 |

We have considered the gravitation of not only the sun, but also the Jupiter and Saturn.

5. Taking into account the aberration, the proper direction of the planet $p_2$ in geocentric inertial reference system (i.e. GCRS) which moves with velocity $\mathbf{V}$ with respect to barycentric reference system (i.e. ICRS) is

$$\mathbf{p}_2 = \left( \beta^{-1} \mathbf{p}_1 + \mathbf{V} + \frac{(\mathbf{p}_1 \cdot \mathbf{V}) \cdot \mathbf{V}}{1 + \beta^{-1}} \right) \Big/ (1 + \mathbf{p}_1 \cdot \mathbf{V}) \tag{7}$$

, where $\mathbf{V} = \dot{\mathbf{E}}_B / c = 0.0057755\,\dot{\mathbf{E}}_B$; $\beta = (1 - \mathbf{V}^2)^{-1/2}$ and the velocity $\mathbf{V}$ is expressed in units of light velocity and equals the velocity of the Earth within accuracy of $\mathbf{V}^2$.

6. The precession and nutation are applied to the proper direction $\mathbf{p}_2$ that converts the position vector of the planet $\mathbf{p}_2$ in GCRS to the vector $\mathbf{p}_3$ with respect to the true equator and equinox of date. First, through the frame bias transformation, from GCRS to J2000.0 mean equator and equinox, then via the precession, to mean equator and equinox of date, finally through the nutation, the rectangular coordinates of position vector in the true equator and equinox are obtained. The resultant apparent direction $\mathbf{p}_3$ are calculated:

$$\mathbf{p}_3 = \mathbf{N}(t)\,\mathbf{P}(t)\,\mathbf{B} \cdot \mathbf{p}_2 \tag{8}$$

, where $\mathbf{p}_3 = \mathbf{r}_{true}(t)$, $\mathbf{p}_2 = \mathbf{r}_{GCRS}(t)$.

The each transformations and their matrixes are showed below:

1) Frame bias

The frame bias is the transformation from ICRS (or GCRS) to the mean equator and equinox of J2000.0. Its matrix is

$$\mathbf{B} = \begin{pmatrix} 1 - \frac{1}{2}(d\alpha_0^2 + \xi_0^2) & d\alpha_0 & -\xi_0 \\ -d\alpha_0 - \eta_0\xi_0 & 1 - \frac{1}{2}(d\alpha_0^2 + \xi_0^2) & -\eta_0 \\ \xi_0 - \eta_0 d\alpha_0 & \eta_0 + \xi_0 d\alpha_0 & 1 - \frac{1}{2}(d\alpha_0^2 + \xi_0^2) \end{pmatrix} \tag{9}$$

, where $d\alpha_0 = -14.6$mas, $\xi_0 = -16.6170$mas, and $\eta_0 = -6.8192$mas are converted to unit of radians. The matrix above is a good approximation to set of rotations $\mathbf{R}_1(-\eta_0)\mathbf{R}_2(\xi_0)\mathbf{R}_3(d\alpha_0)$, where $\mathbf{R}_1$, $\mathbf{R}_2$, $\mathbf{R}_3$ are standard rotations about x, y, z axis.

2) Precession

The precession is the transformation from the mean equator and equinox of the epoch to the mean equator and equinox of date. We consider two precession models: IAU 2000A and IAU 2006.

In IAU2000A precession model the precession matrix is calculated by Capitaine et al. parameterization[5]

$$\mathbf{P}(\chi_A, \omega_A, \psi_A) = \mathbf{R}_3(\chi_A)\,\mathbf{R}_1(-\omega_A)\,\mathbf{R}_3(-\psi_A)\,\mathbf{R}_1(\varepsilon_0) \tag{10}$$

, where $\psi_A$, $\omega_A$, and $\chi_A$, equatorial precession angles, are

$$\begin{aligned}
\psi_A &= 5038.481507T - 1.0790069T^2 - 0.00114045T^3 \\
&\quad + 0.000132851T^4 - 0.0000000951T^5 \\
\omega_A &= \varepsilon_0 - 0.025754T + 0.0512623T^2 - 0.007772503T^3 \\
&\quad - 0.000000467T^4 + 0.0000003337T^5 \\
\chi_A &= 10.556403T - 2.3814292T^2 - 0.001211197T^3 \\
&\quad + 0.000170663T^4 - 0.0000000560T^5
\end{aligned} \tag{11}$$

and $T$ is the number of TDB Julian centuries since 2000 Jan 1, $12^h$ TDB. $\varepsilon_0$ is the mean obliquity of the ecliptic at J2000.0 epoch: $\varepsilon_0 = 843\,81.406$ arcsec.

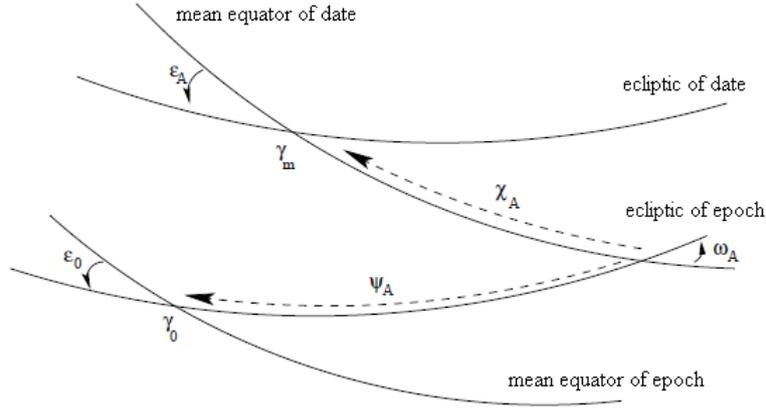

*Figure* 1. Equatorial precession angles $\psi_A$, $\omega_A$, $\chi_A$.

$\gamma_m$ is mean equinox of date, $\gamma_0$ is one of J2000.0 epoch.

In IAU 2006 precession model the Fukushima – Williams parameterization[6] is used. In this parameterization we can get the matrix involving both frame bias and precession:

$$\mathbf{PB} = \mathbf{R}_1(-\varepsilon_A)\, \mathbf{R}_3(-\bar{\psi})\, \mathbf{R}_1(\bar{\phi})\, \mathbf{R}_3(\bar{\gamma}) \tag{12}$$

, where the parameters are

$$\begin{aligned}
\bar{\gamma} &= -0''.052928 + 10''.556378\,T + 0''.4932044\,T^2 - 0''.00031238\,T^3 \\
&\quad - 0''.000002788\,T^4 + 0''.0000000260\,T^5 \\
\bar{\phi} &= +84381''.412819 - 46''.811016\,T + 0''.0511268\,T^2 + 0''.00053289\,T^3 \\
&\quad - 0''.000000440\,T^4 - 0''.0000000176\,T^5 \\
\bar{\psi} &= -0''.041775 + 5038''.481484\,T + 1''.5584175\,T^2 - 0''.00018522\,T^3 \\
&\quad - 0''.000026452\,T^4 - 0''.0000000148\,T^5 \\
\varepsilon_A &= +84381''.406 - 46''.836769\,T - 0''.0001831\,T^2 + 0''.00200340\,T^3 \\
&\quad - 0''.000000576\,T^4 - 0''.0000000434\,T^5
\end{aligned} \tag{13}$$

and its geometrical meaning are expressed in figure 2. *T* is the number of TT Julian centuries since J2000.0.

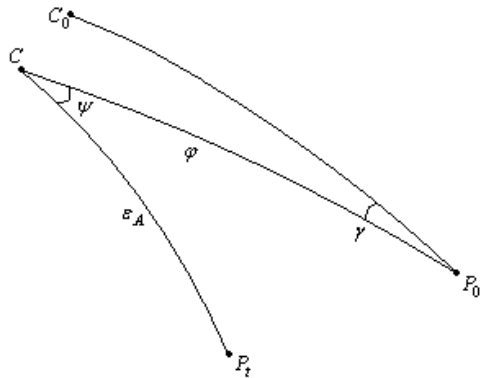

*Figure* 2. Mean pole of date $P_t$, ecliptic pole of date *C* and angles associating mean pole of J2000.0 $P_0$ and ecliptic pole of J2000.0 $C_0$ which are used in Fukushima-Williams parameterization for precession.

(3) Nutation

The nutation is the transformation from the mean equator and equinox of date to the true equator and equinox of date. We consider also two nutation model: IAU 2000A and IAU 2006. Both nutation models give the nutation angles $\Delta\psi$ (in longitude), $\Delta\varepsilon$ (in obliquity). Their meaning are shown in figure 3.

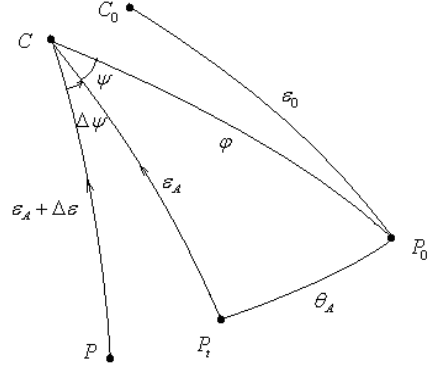

*Figure* 3. Equatorial precession angles $\varepsilon_0$, $\zeta_A$, $\theta_A$, $z_A$, and nutation angles $\Delta\psi$, $\Delta\varepsilon$, where $P_t$ is the mean pole of date, $C$ – ecliptic pole of date, $P_0$ – mean pole of J2000.0, $C_0$ – ecliptic pole of J2000.0, $P$ – true pole of date.

$\Delta\psi$ and $\Delta\varepsilon$ are generally given in the form of long trigonometric series:

$$\Delta\psi = \sum_{i=1}^{N}((S_i + \dot{S}_i T)\sin\Phi_i + C_i'\cos\Phi_i) \qquad (14)$$

$$\Delta\varepsilon = \sum_{i=1}^{N}((C_i + \dot{C}_i T)\cos\Phi_i + S_i'\sin\Phi_i) \qquad (15)$$

, where

$$\Phi_i = \sum_{j=1}^{K} M_{ij}\phi_j(T). \qquad (16)$$

In IAU 2000A model N=1365 and K=14. Fourteen $\phi_j(T)$s are fundamental arguments. These arguments are the orbital angles except for one argument. The main time dependency of the nutation series are involved by these arguments. All coefficients are in unit of arcsec. The tables of M, S(in-phase terms) and C(out-of-phase terms) are published as supplementary of IERS conventions[7] and USNO Circular № 1, 79 [4].

The first 8 fundamental arguments are the mean heliocentric ecliptic longitude of Mercury through Neptune:

$$\begin{aligned}
\phi_1 &= 908103.259872 + 538101628.688982\,T \\
\phi_2 &= 655127.283060 + 210664136.433548\,T \\
\phi_3 &= 361679.244588 + 129597742.283429\,T \\
\phi_4 &= 1279558.798488 + 68905077.493988\,T \\
\phi_5 &= 123665.467464 + 10925660.377991\,T \\
\phi_6 &= 180278.799480 + 4399609.855732\,T \\
\phi_7 &= 1130598.018396 + 1542481.193933\,T \\
\phi_8 &= 1095655.195728 + 786550.320744\,T
\end{aligned} \qquad (17)$$

$T$ is the number of TDB Julian centuries since 2000 Jan 1, 12h TDB. The ninth argument is an approximation to the general precession in longitude:

$$\phi_9 = 5028.8200\, T + 1.112022\, T^2 \tag{18}$$

The last five arguments are the fundamental luni-solar Delaunay arguments: $l$, the mean anomaly of the Moon; $l'$, the mean anomaly of the Sun; $F$, the mean argument of latitude of the Moon; $D$, the mean elongation of the Moon from the Sun, and $\Omega$, the mean longitude of the Moon's mean ascending node:

$$\begin{aligned}
\phi_{10} &= l = 485868.249036 + 1717915923.2178\,T + 31.8792\,T^2 \\
&\quad + 0.051635\,T^3 - 0.00024470\,T^4 \\
\phi_{11} &= l' = 1287104.79305 + 129596581.0481\,T - 0.5532\,T^2 \\
&\quad + 0.000136\,T^3 - 0.00001149\,T^4 \\
\phi_{12} &= F = 335779.526232 + 1739527262.8478\,T - 12.7512\,T^2 \\
&\quad - 0.001037\,T^3 + 0.00000417\,T^4 \\
\phi_{13} &= D = 1072260.70369 + 1602961601.2090\,T - 6.3706\,T^2 \\
&\quad + 0.006593\,T^3 - 0.00003169\,T^4 \\
\phi_{14} &= \Omega = 450160.398036 - 6962890.5431\,T + 7.4722\,T^2 \\
&\quad + 0.007702\,T^3 - 0.00005939\,T^4
\end{aligned} \tag{19}$$

Once the nutation series are evaluated, the nutation matrix can be calculated:

$$\mathbf{N}(t) = \mathbf{R}_1(-\varepsilon)\mathbf{R}_3(-\Delta\psi)\,\mathbf{R}_1(\varepsilon_A) \tag{20}$$

, where $\varepsilon = \varepsilon_A + \Delta\varepsilon$ is the true obliquity.

To be consistent with IAU 2006 precession, the small corrections are added to IAU 2000A nutation[8]:

$$\begin{aligned}
\Delta\psi_{2006} &= \Delta\psi_{2000A} + (0.4697\times 10^{-6} + f)\,\Delta\psi_{2000A} \\
\Delta\varepsilon_{2006} &= \Delta\varepsilon_{2000A} + f\,\Delta\varepsilon_{2000A}
\end{aligned} \tag{21}$$

, where $f \equiv (\dot{J}_2/J_2)T = -2.7774\times 10^{-6}\,T$ and $T$ is the time interval since J2000.0 in Julian centuries (TT).

In the nutation consistent with IAU 2006 model, the above nutation angles are simply added to the corresponding PB angles $\bar{\gamma}, \bar{\phi}, \bar{\psi}, \varepsilon_A$ in (13), then NPB angles $\bar{\gamma}, \bar{\phi}, \psi, \varepsilon$ are gained:

$$\begin{aligned}
\psi &= \bar{\psi} + \Delta\psi_{2006} \\
\varepsilon &= \varepsilon_A + \Delta\varepsilon_{2006}
\end{aligned} \tag{22}$$

These angles are used to compose the NPB matrix based on equinox, which include not only the nutation, but also frame bias and precession:

$$\mathbf{NPB} = \mathbf{R}_1(-\varepsilon)\,\mathbf{R}_3(-\psi)\,\mathbf{R}_1(\bar{\phi})\,\mathbf{R}_3(\bar{\gamma}) \tag{23}$$

The resultant **NPB** can be used to get the vector in the true equator and equinox of date in eq (8).

7. From the rectangular coordinates $\mathbf{p}_3$, the spherical coordinates – right ascension $\alpha$ and declination $\delta$ - are calculated:

$$\mathbf{p}_3 = \mathbf{p}_3(x, y, z), \ \alpha = \arctan\left(\frac{y}{x}\right), \ \delta = \arcsin z \qquad (24)$$

The geocentric distance of the planet for result is the geometric distance between the earth and the planet at the given moment, and it equals the modulus $P$ of $\mathbf{P} = \mathbf{Q}_B(t) - \mathbf{E}_B(t)$ obtained in the first iteration at step 3 that seems to be the case of $\tau = 0$.

**Result**

The apparent positions of planets are the positions with respect to the true equator and equinox of date at geocenter at given moment.

By algorithm stated above, we can calculate the apparent position – apparent right ascension, declination and geocentric distance - of the major planets at $0^h$ TT of date (i.e. moment). The results are compared with "Astronomical Almanac". The table below shows the comparison. The Saturn is chosen to be the sample planet and the fundamental ephemeris DE405 is used. In 2014, TT – UTC = 66.184s is considered.

| Time ($0^h$ TT) | Source | Apparent right ascension | Apparent declination | Geocentric distance(au) |
|---|---|---|---|---|
| Apr, 1,2014 | A.A. | $15^h23^m25^s.817$ | -16°03'27".58 | 9.1295188 |
| | calculation | $15^h23^m25^s.8169503$ | -16°03' 27".5775 | 9.129518827421 |
| June, 1,2014 | A.A. | $15^h06^m53^s.912$ | -14°58' 00".27 | 8.9662247 |
| | calculation | $15^h06^m53^s.9117104$ | -14°58' 00".2679 | 8.966224704421 |

The table shows that within the published uncertainties of A.A. the result we calculated are the same to those in A.A.

**Conclusion**

We composed the calculation system of apparent positions of planets which involves the updated models and conceptions. We used the traditional algorithm in the Astronomical Almanacs and Yearbooks. The JPL fundamental ephemeris DE405/LE405 is the starting point in the calculation procedure. It gives the positions of the planets in ICRS. We considered the gravitational deflections and the aberration et al. to get the proper direction of the planet in GCRS from the position in ICRS. We transformed the position of the planet from GCRS to the true equator and equinox of date that involves taking into account the IAU 2000A/IAU2006 precession-nutation model. The resultant position in the true equator and equinox of date is just the apparent position of the planet we wanted. The calculation results are the same as those in Astronomical Almanac within the published uncertainties.

# References

[1] Mathews, P.M., Herring, T.A., & Buffett B.A., Journal of Geophysical Research, Vol. 107, No. B4, 2068(2002).

[2] Hilton, J. L., Capitaine, N., Chapront, J., Ferrandiz, J. M., Fienga, A., Fukushima, T., Getino, J., Mathews, P., Simon, J. L., Soffel, M., Vondrak, J., Wallace, P., & Williams, J., submitted to Celest. Mech. Dyn. Astr., 94, 351–367(2006).

[3] Capitaine, N., Research in Astron. Astrophys. Vol. 12, No. 8, 1162-1184(2012).

[4] Kaplan, G. H., The IAU Resolutions on Astronomical Reference Systems, Time Scales, and Earth Rotation Models, USNO Circular No. 179(2005).

[5] Capitaine, N., Wallace, P. T., & Chapront, J., Astron. Astrophys., 412, 567(2003).

[6] Fukushima, T., A New Precession Formula, The Astronomical Journal, Vol. 126, 494-534(2003).

[7] IERS Conventions (2003), IERS Technical Note No. 32, eds. McCarthy D. D., & Petit G., (2004).

[8] Wallace, P.T, Capitaine, N., Astron. Astrophys., 459, 981–985(2006).